\documentclass[%
 reprint,
superscriptaddress,
 amsmath,amssymb,
 aps,
prl,
]{revtex4-2}

\usepackage{graphicx}% Include figure files
\usepackage{dcolumn}% Align table columns on decimal point
\usepackage{bm}% bold math
\usepackage{wrapfig}
\usepackage{enumitem}
\usepackage{amsmath,stackengine}
\stackMath
\usepackage{amssymb}
\usepackage{marginnote}
\usepackage{comment}
\usepackage[dvipsnames]{xcolor}
\usepackage[colorlinks, allcolors=NavyBlue]{hyperref}
\usepackage{url}
\usepackage{esint}
\usepackage{multirow}
\usepackage{tikz}
\usetikzlibrary{arrows.meta}

\usepackage{mathtools}
\usepackage{listings}
\usepackage{siunitx}
\usepackage{pgfplots}
\pgfplotsset{compat=1.9, , every axis/.append style={font=\scriptsize}}
\usepackage{xspace}
\usepackage{orcidlink}

\newcommand{\mb}{\mathbf}

\newcommand{\C}{\mathfrak{C}}
\newcommand{\Cel}{\mathfrak{C}^{\text{el}}}
\renewcommand{\u}{\mathfrak{u}}
\renewcommand{\phi}{\varphi}
\newcommand{\E}{\mathcal{E}}

\begin{document}

\title{Ubiquity of Uncertainty in Neuron Systems}

\author{Brandon B. Le\,\orcidlink{0009-0002-7354-9136}}
\email[Contact author: ]{sxh3qf@virginia.edu}
\affiliation{
Department of Physics, University of Virginia, Charlottesville, Virginia 22904, USA
}
\affiliation{
Department of Mathematics, University of Virginia, Charlottesville, Virginia 22904, USA
}
\author{Bennett Lamb\,\orcidlink{0009-0008-5067-1458}}
\email{xkb2jd@virginia.edu}
\affiliation{
Department of Physics, University of Virginia, Charlottesville, Virginia 22904, USA
}
\affiliation{
Department of Mathematics, University of Virginia, Charlottesville, Virginia 22904, USA
}
\author{Luke Benfer}
\email{rdd3cs@virginia.edu}
\affiliation{
Department of Mathematics, University of Virginia, Charlottesville, Virginia 22904, USA
}
\author{Sriharsha Sambangi}
\email{agr6bq@virignia.edu}
\affiliation{
Department of Neuroscience, University of Virginia, Charlottesville, Virginia 22908, USA
}
\author{Nisal Geemal Vismith}
\email{gsx3qz@virginia.edu}
\affiliation{
Department of Physics, University of Virginia, Charlottesville, Virginia 22904, USA
}
\author{Akshaj Jagarapu\,\orcidlink{0009-0007-4767-4431}}
\email{akshaj@uchicago.edu}
\affiliation{
Department of Mathematics, University of Chicago, Chicago, Illinois 60637, USA
}

\date{\today}% It is always \today, today,1

\begin{abstract}
We demonstrate that final-state uncertainty is ubiquitous in multistable systems of coupled neuronal maps, meaning that predicting whether one such system will eventually be chaotic or nonchaotic is often nearly impossible. We propose a ``chance synchronization'' mechanism that governs the emergence of unpredictability in neuron systems and support it by using basin classification, uncertainty exponent, and basin entropy techniques to analyze five simple discrete-time systems, each consisting of a different neuron model. Our results illustrate that uncertainty in neuron systems is not just a product of noise or high-dimensional complexity; it is also a fundamental property of low-dimensional, deterministic models, which has profound implications for understanding brain function, modeling cognition, and interpreting unpredictability in general multistable systems.
\end{abstract}

\maketitle

%\tableofcontents

\begin{comment}
    NOVELTY:
    One of the first studies to find qualitative, extreme final state uncertainty in neuron systems.
    The first study to apply basin entropy techniques to analyze neuron systems.
\end{comment}

\textit{Introduction}---Dynamical systems with multiple coexisting attractors can exhibit final-state uncertainty \cite{grebogi-final-state}, a remarkable property that poses a significant barrier to predictability \footnote{In the literature, the terms final-state uncertainty, final-state sensitivity, and final-state unpredictability all refer to the same phenomenon and can be used interchangeably.}. Specifically, a multistable system with final-state uncertainty is one whose eventual state is highly sensitive to its initial conditions, which manifests geometrically as a fractal basin boundary \cite{mcdonald}. In even more extreme cases, basins may be described as riddled \cite{alexander}, where every point in a basin is arbitrarily close to a point belonging to another basin. The existence of fractal basin boundaries or riddled basins in a system has a profound impact on our ability to model, predict, and control it, especially in the context of neuronal systems.

The mathematical modeling of neurons has been a longstanding goal in neuroscience. Since the pioneering work of Hodgkin and Huxley \cite{hh}, many models have been developed in an attempt to capture the complex dynamics of biological neurons \cite{chay, buchholtz, izhikevich-model, hindmarsh, courbage}. Indeed, while the dynamics of neuron systems have been extensively studied \cite{ibarz, rabinovich}, comparatively little research has focused on their geometrical properties and the possible presence of final-state uncertainty. Among the limited examples of neuronal systems discovered to exhibit final-state uncertainty are a network of theta neurons with three periodic orbit attractors \cite{so}, an adaptive synapse-based neuron model with up to twelve heterogeneous attractors \cite{bao-neuron}, and a network of Izhikevich neurons with synchronized and unsynchronized attractors \cite{aristides}. All of these systems are complex continuous-time models, but recently, a system composed of only two coupled discrete-time neurons was also found to exhibit final-state uncertainty \cite{le}. Importantly, rather than having multiple qualitatively similar attractors---such as distinct periodic orbits---this system contains coexisting nonchaotic and chaotic attractors. This suggests that the discrete-time nature of neuronal maps facilitates the emergence of qualitative final-state uncertainty even in simple neuron systems.

In this Letter, we show that this qualitative final-state unpredictability is, in fact, ubiquitous in simple systems of coupled neuronal maps containing coexisting nonchaotic and chaotic attractors. To do this, we use the methods of basin classification \cite{sprott}, uncertainty exponents \cite{mcdonald}, and basin entropy \cite{basin-entropy} to analyze the geometrical properties of a broad set of systems featuring neuron models and coupling schemes that span a wide range of dynamical behaviors, abstraction levels, and complexity. We discuss the mechanism behind the universality of this phenomenon and the neurobiological implications of neuron systems being fundamentally uncertain.

\textit{The models}---We consider five discrete-time models (Models~1-5), each containing a small number of coupled low-dimensional neurons. Model~1 is the system explored in Ref. \cite{le}: two asymetrically electrically coupled nonchaotic Rulkov neurons \cite{rulkov}, which we use as a benchmark model. In this four-dimensional system, the $i$th neuron's dynamics are given by the iteration function
\begin{subequations}
\label{eq:model1}
\begin{equation}
    \begin{dcases}  
        \begin{aligned}
            x_i(k+1) &= f(x_{i}(k), y_{i}(k)+\Cel_i(k); \alpha) \\
            y_i(k+1) &= y_{i}(k) - \mu x_{i}(k) + \mu[\sigma + \Cel_i(k)] \\
        \end{aligned}
    \end{dcases}, \label{eq:model1a}
\end{equation}
where the fast variable function $f$ is defined as
\begin{equation}
    f(x,y;\alpha) = 
    \begin{cases}
        \alpha/(1-x) + y, & x\leq 0 \\
        \alpha + y, & 0 < x < \alpha + y \\
        -1, & x\geq \alpha + y
    \end{cases}. \label{eq:model1b}
\end{equation}
\end{subequations}
Here, $k$ represents the discrete time step, $x_i$ is the fast variable (representing the voltage) of the $i$th neuron, $y_i$ is the slow variable of the $i$th neuron, $\Cel_1=g_1(x_2 - x_1)$ and $\Cel_2=g_2(x_1-x_2)$ are electrical coupling terms representing a flow of current, $g_i$ is the electrical coupling strength (conductance) associated with the $i$th neuron, $\alpha$ and $\sigma$ are parameters, and $0<\mu\ll 1$ is a small parameter to make $y$ slow. Model~2 is composed of two asymmetrically electrically coupled Chialvo neurons \cite{chialvo}, a four-dimensional model constructed based on Model~1. The $i$th neuron has the iteration function
\begin{equation}
    \begin{dcases}  
        \begin{aligned}
            x_i(k+1) &= x_i(k)^2 e^{y_i(k)-x_i(k)} + I + \Cel_i(k) \\
            y_i(k+1) &= ay_i(k) - bx_i(k) + c \\
        \end{aligned}
    \end{dcases},
    \label{eq:model2}
\end{equation}
where $x_i$ is the voltage variable, $y_i$ is the recovery variable, $a$, $b$, and $c$ are recovery parameters, $I$ models the injection of direct current, and the electrical coupling terms $\Cel_i(k)$ are defined in the same way as Model~1.

In Model~3, we explore nonuniform pulse coupling \cite{pulse-coupled-ns} with a system of three Nagumo-Sato neurons \cite{nagumo-sato}, each of which are characterized by one voltage variable $x_i$:
\begin{equation}
    x_i(k+1) = x_i(k)/b + a - H(x_i(k)) + \C_i^{\text{pul}}(k).
    \label{eq:model3}
\end{equation}
Here, $a$ and $b$ are parameters, $H(\cdot)$ is the Heaviside step function, $\C_i^{\text{pul}}(k) = \sum_{j\neq i}\kappa_jH(x_j(k))$ is the pulse coupling term, and $\kappa_j$ is the pulse coupling strength of the $j$th neuron, which represents the amplitude of the spike neuron $j$ sends to the other neurons when it fires. Model~4 examines the uniform chemical synaptic coupling \cite{aristides} of three (discrete-time) Izhikevich neurons \cite{izh-map} defined by the six-dimensional iteration function
\begin{subequations}
\label{eq:model4}
\begin{equation}
    \begin{dcases}  
        \begin{aligned}
            x_i(k+1) &= 0.04x_i(k)^2 + 6x_i(k) + 140 \\
             &\mathrel{\phantom{=}} -\, y_i(k) + I + \C_i^{\text{ch}}(k) \\
            y_i(k+1) &= 0.004x_i(k) + 0.98y_i(k)
        \end{aligned}
    \end{dcases}
    \label{eq:model4a}
\end{equation}
for $x_i(k)<30$, with resetting mechanism
\begin{equation}
    \begin{dcases}  
        \begin{aligned}
            x_i(k+1) &= c \\
            y_i(k+1) &= y_i(k) + d
        \end{aligned}
    \end{dcases}
    \label{eq:model4b}
\end{equation}
\end{subequations}
for $x_i(k)\geq 30$. Here, $x_i$ is the fast (voltage) variable, $y_i$ is the slow (recovery) variable, $c$, $d$, and $I$ are parameters, and $\C_i^{\text{ch}} = x_i\sum_{j\neq i}\gamma (1+e^{-7x_j})^{-1}$ is the chemical coupling term, where $\gamma$ is the chemical coupling strength.

To demonstrate that uncertainty exists beyond simple systems of identical neurons, we introduce Model~5, a more complex heterogeneous neuronal system from Ref. \cite{luo} consisting of a chaotic Rulkov neuron ($x_1$ and $y_1$) \cite{rulkov2}, a FitzHugh-Nagumo neuron ($x_2$ and $y_2$) \cite{fh}, and a Hindmarsh-Rose neuron ($x_3$ and $y_3$) \cite{hindmarsh-rose} coupled via a memristor \cite{memristor}:
\begin{equation}
    \begin{dcases}  
        \begin{aligned}
            x_1(k+1) &= 4.5/[1+x_1(k)^2] + y_1(k) - \C^{\text{mem}}(k) \\
            y_1(k+1) &= y_1(k) - 0.5x_1(k) - 0.55 \\
            x_2(k+1) &= x_2(k) + 0.6[x_2(k) - x_2(k)^3/3 - y_2(k) \\
            &\mathrel{\phantom{=}} +\, \C^{\text{mem}}(k) + \Cel(k)] \\
            y_2(k+1) &= y_2(k) + 1.8[x_2(k) + 0.5 - 0.9y_2(k)] \\
            x_3(k+1) &= x_3(k) + 0.45[y_3(k) - 0.2x_3(k)^3 \\
            &\mathrel{\phantom{=}} +\, 0.6x_3(k)^2 - \Cel(k)] \\
            y_3(k+1) &= y_3(k) + 0.45[0.2-0.1x_3(k)^2 - y_3(k)] \\
            \phi(k+1) &= -0.1\phi(k)^3 + 1.1\phi(k) + 0.1[x_1(k)-x_2(k)]
        \end{aligned}
    \end{dcases},
    \label{eq:model5}
\end{equation}
where $\phi$ is the internal flux of the memristor, $\C^{\text{mem}} = \xi(x_1-x_2)(0.3\phi-0.5\tanh\phi)$ and $\Cel = g(x_3-x_2)$ are the memristor and electrical coupling terms, respectively, and $\xi$ and $g$ are coupling parameters.

In all of these models, we choose biologically plausible parameter values that result in nonchaotic dynamics for individual neurons \cite{ibarz}. A summary of the models and a list of the parameters used can be found in Table \ref{tab:data}.

\begin{table*}[t]
    \centering
    \caption{\label{tab:data} State space dimensions ($n$), parameter values, coexisting attractors, basin classifications (BC), uncertainty exponents ($\u$), and basin entropy ($S_b$) regressions for Models~1-5. The specified region of state space $\Omega$ is used for $\u$ and $S_b$ calculations, the $\u$ values are calculated from a $\varrho(\epsilon)$ regression, and the $S_b$ regressions take the approximate form $\ln(S_b) = \u\ln(\epsilon) + \ln\left(s\ln(N_A)\right)$ if $p_{i,j}\approx 1/m_i\,\forall j$. These results show that all five models exhibit final-state uncertainty ($\u<1$) and fractal basin boundaries ($d=n-\u>n-1$), as well as an approximate agreement of the $S_b$ regressions with the $\u$ values.}
    \begin{ruledtabular}
    \begin{tabular}{cccccccc}
        Model~& $n$ & Parameters & Attractors & BC & $\Omega$ & $\u$ & $S_b$ reg. \\
        \hline
        \multirow{2}{*}{\ref{eq:model1} (Rulkov)} & \multirow{2}{*}{4} & $\sigma=-0.5, \alpha=4.5,$ & Chaotic & 2 & $([-2,2]$ & \multirow{2}{*}{0.04} & $\ln(S_b) = $ \\ \cline{4-5}
         & & $g_1=0.05, g_2=0.25$ & Nonchaotic & 2 & $\times[-1,5])^2$ & & $0.03\ln(\epsilon) - 1.52$ \\
        \hline
        \multirow{2}{*}{\ref{eq:model2} (Chialvo)} & \multirow{2}{*}{4} & $a=1.0, b=2.2, c=0.26,$ & Chaotic & 1 & $([-4,4]$ & \multirow{2}{*}{0.13} & $\ln(S_b) = $ \\ \cline{4-5}
         & & $I=0.04, g_1=0.05, g_2=0.3$ & Nonchaotic & 3 & $\times [-4,4])^2$ & & $0.12\ln(\epsilon) - 0.96$ \\
        \hline
        \multirow{2}{*}{\ref{eq:model3} (N-S)} & \multirow{2}{*}{3} & $a=0.18, b=1.15, $ & Unsynchronized & 2 & \multirow{2}{*}{$[-1,2]^3$} & \multirow{2}{*}{0.45} & $\ln(S_b) = $ \\ \cline{4-5}
         & & $\kappa_1=0.005, \kappa_2=0.01, \kappa_3=0.02$ & Synchronized & 2 & & & $0.45\ln(\epsilon) - 1.19$ \\
        \hline
        \multirow{2}{*}{\ref{eq:model4} (Izhikevich)} & \multirow{2}{*}{6} & $c=-55, d=8$, & Chaotic & 2 & $([-200,30]$ & \multirow{2}{*}{0.03} & $\ln(S_b) = $ \\ \cline{4-5}
         & & $I=15$, $\gamma = 0.5$ & Nonchaotic & 2 & $\times [-50,30])^3$ & & $0.04\ln(\epsilon) - 0.52$ \\
        \hline
        \multirow{2}{*}{\ref{eq:model5} (Hetero)} & \multirow{2}{*}{7} & \multirow{2}{*}{$\xi=-0.2, g=0.4$} & Chaotic & 4 & $([-2,2]\times[-1,1])^3$ & \multirow{2}{*}{0.23} & $\ln(S_b) = $ \\ \cline{4-5}
         & & & Nonchaotic & 4 & $\times[0,4]$ & & $0.23\ln(\epsilon) - 0.41$
    \end{tabular}
    \end{ruledtabular}
\end{table*}

\textit{Methodology}---To characterize the basins of attraction of these neuron systems, we use the classification method of Sprott and Xiong \cite{sprott}, which we briefly outline here. Consider an $n$-dimensional dynamical system with attractor $A$ and associated basin $\hat{A}$. Define $\xi=(\mathbf{x} - \langle A\rangle)/\sigma_A$ to be the ``normalized distance'' of a state $\mathbf{x}$ from $A$, where $\langle A\rangle$ is the ``center of mass'' of $A$ and $\sigma_A$ is the standard deviation of $A$. Then, define $P(\xi)$ to be the probability that an initial state $\mathbf{x}_0$, selected at random from the $n$-ball with radius $\xi$ centered at $\langle A\rangle$, is in the basin $\hat{A}$. In the limit $\xi\to\infty$, $P(\xi)$ usually follows a power law $P(\xi) = P_0 / \xi^\gamma$. The basin $\hat{A}$ is then classified based on these parameters $P_0$ and $\gamma$: Class~1 basins take up all of space barring a set of finite measure ($P_0=1$, $\gamma=0$), Class~2 basins occupy a fixed fraction of state space ($P_0<1$, $\gamma=0$), Class~3 basins extend to infinity but occupy increasingly small fractions of state space further out ($0<\gamma<n$), and Class~4 basins occupy a finite region of state space and have a well-defined relative size $\xi_0 = P_0^{1/n}$ ($\gamma = n$). In practice, we compute $P(2^m)$ using a Monte Carlo method and iteratively compute $P(2^{m+1})$ using the shell method derived in Ref. \cite{bn}.

Uncertainty exponents \cite{mcdonald} quantify final-state uncertainty by focusing on the fractalization of the basin boundary. Specifically, let $\mathbf{x}_0\in\hat{A}$ be a randomly chosen initial state in a given region of state space $\Omega$. If we introduce a small uncertainty $\epsilon$ to $\mathbf{x}_0$, there is a probability $\varrho(\epsilon)$ that the perturbed initial state will not be attracted to $A$. In the limit $\epsilon\to 0$, $\varrho(\epsilon)$ follows a power law $\varrho(\epsilon)\sim \epsilon^{\mathfrak{u}}$, where $\mathfrak{u}$ is the uncertainty exponent. An uncertainty exponent less than 1 indicates that the system exhibits final-state uncertainty, and the fractal dimension $d$ of the basin boundary is related to the uncertainty exponent by $d = n-\mathfrak{u}$. Therefore, the smaller the uncertainty exponent, the more extreme the final-state uncertainty and the more fractal the basin boundary. In the limit, a riddled basin has $\mathfrak{u}=0$ because, by definition, lowering the initial-state uncertainty $\epsilon$ does not improve the final-state uncertainty \cite{alexander}.

Basin entropy \cite{basin-entropy} is another method of quantifying uncertainty that encapsulates more information than either uncertainty exponents or basin stability \cite{basin-stability} alone. To compute it for a system with $N_A$ attractors, we define a ``color'' function that labels an initial state $\mb{x}_0\in\Omega$ with an integer from 1 to $N_A$ according to which basin it belongs to. Covering $\Omega$ with $n$-dimensional boxes of side length $\epsilon$, we randomly sample points from each box $i$ and assign to each of them a color $j$. Denoting $p_{i,j}$ to be the probability that a point chosen at random from box $i$ has color $j$ \footnote{In practice, we compute this probability by sampling 25 initial states per box, which is standard.}, $m_i\in\{1,\hdots,N_A\}$ to be the number of colors inside the box $i$, and $N$ to be the total number of boxes, the basin entropy $S_b$ is defined as
\begin{equation}
    S_b = \frac{1}{N}\sum_{i=1}^N\sum_{j=1}^{m_i}p_{i,j}\ln(1/p_{i,j}),
    \label{eq:S_b}
\end{equation}
which ranges from $0$ (only one attractor) to $\ln(N_A)$ (completely randomized basins). An interpretation of basin entropy arises by assuming that the colors inside a given box $i$ are equiprobable, i.e., $p_{i,j} = 1/m_i$ for all $j$, and that there is only one boundary between all $N_A$ basins. Then, we can write Eq.~\eqref{eq:S_b} as $S_b = s\ln(N_A)\epsilon^{\u}$, where $s$ is a constant corresponding to the size of the boundary. This is a power law relationship that allows us to easily analyze the effects of $s$, $N_A$, and $\u$ on $S_b$. 

\begin{comment}
A natural extension to basin entropy is boundary basin entropy $S_{bb} = S/N_b$, where $N_b$ is the number of boxes containing more than one color. While $S_b$ is sensitive to the size of basins, $S_{bb}$ is not. It does, however, provide a sufficient condition to determine whether a basin boundary is fractal, namely, $S_{bb}>\ln2\implies\u<1$ \cite{basin-entropy}.
\end{comment}

\begin{figure*}
    \centering
    \includegraphics[width=\textwidth]{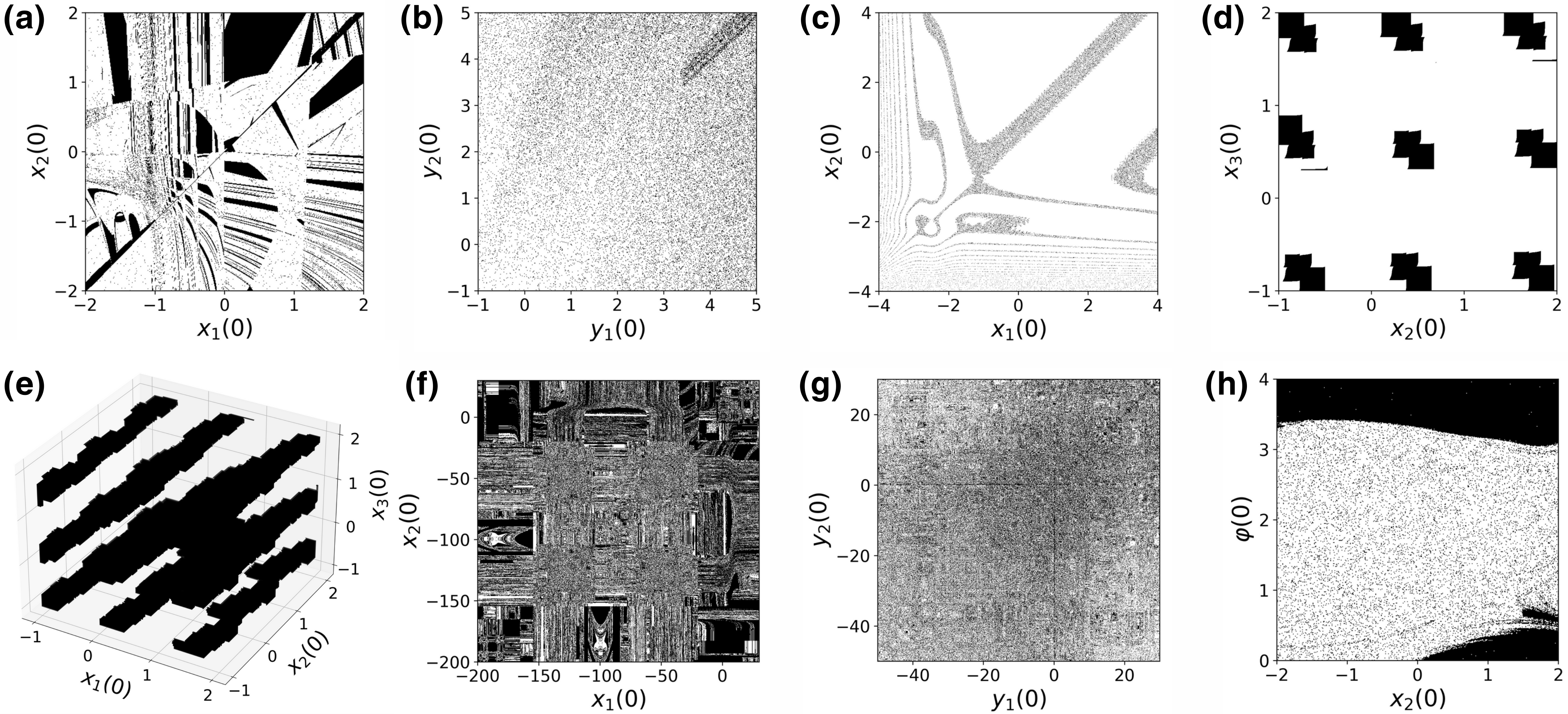}
    \caption{Basins of attraction corresponding to the chaotic/unsynchronized (white) and nonchaotic/synchronized (black) attractors in the $\Omega$ regions of Models~1–5, with parameters and $\Omega$ specified in Table \ref{tab:data}. (a) Model~1, $(x_1,-3.25,x_2,-3.25)$ slice; (b) Model~1, $(-1,y_1,1,y_2)$ slice; (c) Model~2, $(x_1,-2,x_2,-3)$ slice; (d) Model~3, $(0.4,x_2,x_3)$ slice; (e) Model~3, full $\Omega$; (f) Model~4, $(x_1, 0, x_2, 0, 0, 0)$ slice; (g) Model~4, $(0, y_1, 0, y_2, 0, 0)$ slice; (h) Model~5, $(0, 0, x_2, 0, 0, 0, \varphi)$ slice.}
    \label{fig:basins}
\end{figure*}

\textit{Results and discussion}---We begin by establishing the existence of coexisting, qualitatively different attractors in the models with parameters specified in Table \ref{tab:data}. For all of the models except Model~3, attractor orbits are characterized using the maximal Lyapunov exponent $\lambda_1$, which we compute using the standard QR factorization method for computing Lyapunov spectra \cite{eckmann} on orbits of sufficient length $T$ for convergence. This reveals two qualitatively different attractors in the coupled neuron systems: chaotic ($\lambda_1 > 0$) and nonchaotic ($\lambda_1 < 0$). On the other hand, because of the simplicity of Model~3 (i.e., piecewise linear and no recovery/slow variables), we characterize its orbits using a synchronization error
\begin{equation}
    \E = \frac{1}{T-T_0}\sum_{k=T_0}^{T}\sum_{i=1}^{3}|x_i(k) - \overline{x}(k)|,
\end{equation}
where $T$ is the number of timesteps in the orbit, $T_0$ is the ``burn-in time,'' and $\overline{x}(k)$ is the average voltage of the three neurons at timestep $k$. Using a synchronization cutoff of $\E_0 = 0.2$ \footnote{This choice of synchronization cutoff is reasonable because we find that no orbits with synchronization error between $0.15<\E<0.22$. Therefore, it makes sense to call orbits with $\E<0.2$ synchronized, and vice versa.}, we designate orbits into two qualitatively different attractors: unsynchronized ($\E>\E_0$) and synchronized ($\E<\E_0$).

We now describe the general ``chance synchronization'' mechanism that permits qualitatively different attractors and extreme final-state uncertainty to emerge in Models~1-5, as well as in many other neuronal systems. When coupled neurons are individually nonchaotic and exhibit similar individual dynamics (as in this study), it is clear that if the initial conditions of each neuron are identical or very close, their dynamics will synchronize, resulting in a convergence to the nonchaotic attractor. However, when initial conditions differ across neurons, their individual dynamics misalign, so the coupling between them---whether it be electrical, chemical, pulse, or memristor-based---will cause the neurons to interact with each other. The complexity of the individual dynamics and coupling connections often results in these interactions yielding chaotic dynamics, bringing the orbit to a chaotic attractor. This tendency is further amplified in discrete-time systems, which allow multiple neurons to readily fall into complex frequency ratios, resulting in persistent, long-term dynamics on the chaotic attractor. Final-state uncertainty arises because different neurons having different initial states does not always result in chaotic dynamics. Sometimes, certain initial conditions will result in the neurons synchronizing with each other by chance; for example, if the initial conditions happen to be such that the coupling brings two neurons to a similar point in their periodic cycle, then the neurons will lock onto each other and synchronize. As a result, the system becomes highly sensitive to initial conditions in terms of which attractor it ultimately approaches---the definition of final-state uncertainty.

By means of this chance synchronization mechanism, final-state uncertainty emerges in all of our models. Table \ref{tab:data} shows that the uncertainty exponent $\mathfrak{u}$ in the prescribed $\Omega$ region is less than 1 for all five systems, and Fig.~\ref{fig:basins} presents a visualization of the systems’ basins and fractal basin boundaries characteristic of final-state uncertainty, where black points belong to the basin of the nonchaotic/synchronized attractor and white points belong to the chaotic/unsynchronized basin. Among the five systems, we observe different types of fractal basin boundaries with varying levels of uncertainty, from solid regions with jagged, rough edges [e.g., Fig.~\ref{fig:basins}(e) with $\u=0.45$] to completely intermingled black and white points [e.g., Fig.~\ref{fig:basins}(b) with $\u=0.04$]. 

We first discuss the most extreme cases of final-state uncertainty: Model~1 [Fig.~\ref{fig:basins}(b)] and Model~4 [Fig.~\ref{fig:basins}(g)], whose basins are nearly riddled with $\u\approx 0$. Using the power law interpretation of $\u$, we find that to reduce uncertainty in the final state by a factor of 10, we need to reduce uncertainty in the initial state by on the order of $10^{25}$ for Model~1 and $10^{33}$ for Model~4. The reason that final-state uncertainty is pushed to the extreme in these two models is due to their slow-fast mechanism [see Eqs.~\eqref{eq:model1a} and \eqref{eq:model4a}]. Specifically, consider an initial state in $\Omega$. Since the voltage variables $x$ evolve so much faster than the slow variables $y$, once the $y$ variables converge to values near the system's attractors, the $x$ variables will have undergone such comparatively faster evolution that they are essentially randomized. This justifies the system's near riddled basins and the appearance of seemingly random distributions of white and black points in Figs.~\ref{fig:basins}(b) and \ref{fig:basins}(g), which show the $(y_1,y_2)$ slices of $\Omega$. However, the basins of these slow-fast systems are not completely riddled with $\u=0$ because the $y$ variables are not infinitely slow relative to $x$; thus, initial conditions that are sufficiently close will still evolve closely enough to be attracted to the same final state once the $y$ variables approach the vicinity of the attractors.

Next, we compare the basin classification, uncertainty exponent, and basin entropy results in Table \ref{tab:data} across the five models to highlight how these methods can enhance understanding of the geometrical and final-state uncertainty properties of neuron systems. The basin classifications provide us with a picture of what the basins look like outside of the $\Omega$ regions displayed in Fig.~\ref{fig:basins}. Specifically, since Models~1, 3, and 4 have Class~2 basins, the fraction of states that converge to either attractor remains roughly the same regardless of how far away the initial states are from the attractors. This suggests that the amount of uncertainty is relatively independent of the magnitude of the initial states. On the other hand, since Model~2 has a Class~3 nonchaotic basin, the farther an initial state is from its attractor, the lower chance there is for the neurons to immediately synchronize. Therefore, in Model~2, uncertainty is lowered when considering neurons with initial voltage and recovery variables of high magnitude. Finally, Model~5 having Class~4 basins reflects the fact that the model is only stable for a finite region of initial states, which $\Omega$ lies inside of. Outside this region, the model fails and the initial state does not converge to either attractor.

Considering the final-state uncertainty in terms of the fractal basin boundary $d=n-\mathfrak{u}$, we observe that the uncertainty is negatively correlated with the abstraction of the biological neuron model \footnote{The levels of abstraction of the neuron models are summarized in Ref. \cite{ibarz}}. Namely, among the homogeneous neuron models, the highly abstract Nagumo-Sato model exhibits the lowest uncertainty, the moderately abstract Chialvo model shows intermediate uncertainty, and the minimally abstract Rulkov and Izhikevich models have the highest uncertainty. This trend strongly suggests that extreme final-state uncertainty emerges in real biological neurons. Considering the final-state uncertainty in terms of the basin entropy $S_b$ regression, as expected, there is an agreement between the slope of the regression and the uncertainty exponent $\u$. Another important observation is that although $\u$ indicates that the uncertainties in Models~1 and 4 are similar, the $S_b$ regression indicates that the uncertainty in Model~4 [$S_b(\epsilon=0) \approx 0.6$] is much higher than that of Model~1 [$S_b(\epsilon=0) \approx 0.2$], which reflects the basin visualizations in Figs.~\ref{fig:basins}(b) and \ref{fig:basins}(g). Specifically, in Model~1, most orbits are chaotic ($\approx 85\%$ of $\Omega$) because it is unlikely for the neurons to synchronize by chance, but in Model~4, due to the higher coupling strength ($\gamma = 0.5$ vs. $g=0.05,0.25$) leading to more frequent synchronization, there is a more equal mix of chaotic ($\approx 69\%$) and nonchaotic orbits ($\approx 31\%$). This demonstrates how basin entropy can capture important final-state uncertainty properties beyond boundary fractalization.

%higher uncertainty from lower abstraction (ibarz)
%Chaotic basin size: 1/85%, 3/69%

\textit{Conclusions and outlook}---In this Letter, we detailed a mechanism of chance synchronization that leads to a qualitative final-state uncertainty emerging from sensitive dependence on initial conditions. Using five different models spanning a range of biophysical complexity, dynamical behavior, and coupling schemes, we demonstrate that this uncertainty is ubiquitous in neuron systems and is amplified by discretized time, different timescales, and biological realism. We also highlight the utility of basin classification, uncertainty exponents, and basin entropy in characterizing the final-state uncertainty of neuron systems. Our results indicate that even in simple, deterministic neuron systems, predicting whether the system will eventually behave chaotically or nonchaotically is often extremely difficult. We conclude that many neuron systems are fundamentally unpredictable.

This paradigm shift carries profound implications across neuroscience, nonlinear science, and beyond. For example, intrinsic uncertainty in neural dynamics may underlie the natural variability observed in perception \cite{perception}, behavior \cite{behavior}, decision-making \cite{decision-making}, and memory retrieval \cite{memory-retrieval}, shedding light on why even subtle differences in neural states can produce drastically different outcomes. In neurological disorders such as Alzheimer’s disease \cite{alzheimer, alzheimer-2}, Parkinson’s disease \cite{parkinson, parkinson-2}, and epilepsy \cite{epilepsy-1, epilepsy-2, epilepsy-3}, where typical neural patterns break down, this inherent unpredictability could be exacerbated or dysregulated, potentially driving symptom progression, seizure generation, or clinical variability. More broadly, embracing unpredictability as a fundamental aspect of neural computation challenges longstanding assumptions in cognitive modeling, prompting a critical reevaluation of large-scale brain models that often depend on stable, noise-driven dynamics \cite{cognitive-1, cognitive-2}. Beyond biological applications, our tripartite methodology and chance synchronization mechanism are also applicable to analyzing the unpredictability of nonlinear systems in other fields, such as climate systems \cite{wunderling, margazoglou}, celestial mechanics \cite{deAssis, valade}, laser physics and nonlinear optics \cite{peil, flunkert, meucci, mousavi}, chemical reaction networks \cite{tang, nicolaou}, and agent-based systems \cite{mendes, bertolotti, lee-sync}.

Future work will analyze each of these systems in greater depth, further exploring the final-state sensitivity properties emerging from a more careful treatment of the models' many complex dynamical regimes \cite{osipov, wang2, ramirez-avila, lopez, ring, n-dim, jampa, used, ramirez, kuznetsov, wang-chialvo, nakagawa, oku, ito, izhikevich-article, izhikevich-polychronization, shanahan, wang-memristor}. We will also explore the existence of the Wada property in neuron systems, an even more extreme form of final-state uncertainty that arises when three or more basins share the same boundary \cite{kennedy}. Finally, in a future paper, we will go beyond the phenomenological models explored in this work by examining the existence of final-state uncertainty in more biophysically grounded continuous-time neuron systems, bringing us closer to linking our theoretical results with electrophysiological data. We therefore recommend that work be conducted to experimentally demonstrate the existence of final-state uncertainty in real biological neuron systems. \\

%\cite{osipov, wang2, ramirez-avila, lopez, n-dim, jampa, used, ramirez, kuznetsov, wang-chialvo, nakagawa, oku, ito, izhikevich-article, izhikevich-polychronization, shanahan, wang-memristor}

\textit{Acknowledgements}---B.B.L. is grateful to Grace T. Bai and Meilin Ranjan for beneficial discussions.

\bibliography{refs_uncert_title}% Produces the bibliography via BibTeX.

\end{document}